\documentclass[sigconf]{acmart}

\acmConference[NLBSE 2024]{3rd International Workshop on Natural Language-based Software Engineering}{April 2024}{Lisbon, Portugal}

\usepackage{chngcntr}

\usepackage{graphicx}

\usepackage{amsmath}

\usepackage{subcaption}
\usepackage{amsfonts}
\usepackage{algorithmic}
\usepackage{textcomp}
\usepackage{xcolor}
\usepackage{verbatim}
\usepackage{tabularx}
\usepackage{xspace}
\usepackage{booktabs}
\usepackage{soul}
\usepackage{url}
\usepackage{enumitem}
\usepackage{colortbl}
\usepackage{multirow, multicol, makecell}
\usepackage{url} 
\usepackage{xspace}
\usepackage{longtable}
\usepackage{afterpage}
\usepackage{framed}
\usepackage{natbib}
\usepackage{array}
\usepackage{float}
\usepackage{kvoptions}
\title{Emotion Classification In Software Engineering Texts: A Comparative Analysis of Pre-trained Transformers Language Models}

\author{Mia Mohammad Imran}
\affiliation{\institution{Virginia Commonwealth University}
  \city{Richmond, Virginia}
  \country{USA}
}
\email{imranm3@vcu.edu}

\begin{abstract}
Emotion recognition in software engineering texts is critical for understanding developer expressions and improving collaboration. This paper presents a comparative analysis of state-of-the-art Pre-trained Language Models (PTMs) for fine-grained emotion classification on two benchmark datasets from GitHub and Stack Overflow. We evaluate six transformer models - BERT, RoBERTa, ALBERT, DeBERTa, CodeBERT and GraphCodeBERT against the current best-performing tool SEntiMoji. Our analysis reveals consistent improvements ranging from 1.17\% to 16.79\% in terms of macro-averaged and micro-averaged F1 scores, with general domain models outperforming specialized ones. To further enhance PTMs, we incorporate polarity features in attention layer during training, demonstrating additional average gains of  1.0\% to 10.23\% over baseline PTMs approaches. Our work provides strong evidence for the advancements afforded by PTMs in recognizing nuanced emotions like Anger, Love, Fear, Joy, Sadness, and Surprise in software engineering contexts. Through comprehensive benchmarking and error analysis, we also outline scope for improvements to address contextual gaps.
\end{abstract}

\begin{document}

\maketitle
\begin{sloppypar}

\section{INTRODUCTION}

The exploration of emotion classification in software engineering (SE) texts has garnered considerable attention in recent years~\cite{novielli2023emotion, lin2022opinion}, leading to the development of various emotion classification tools by researchers. Notable among these are ESEM-E by Murgia et al.~\cite{esem-e}, EMTk by Calefato et al.~\cite{emtk}, and SEntiMoji by Chen et al.~\cite{sentimoji}. Chen et al.'s comparative study, utilizing emotion datasets extracted from Stack Overflow and JIRA~\cite{ortu2016emotional, novielli2018gold}, affirmed the superior performance of SEntiMoji compared to EMTk and ESEM-E, solidifying its prominence in the domain. However, compared to open-domain, Pre-Trained Language Models (PTMs) rarely have been used in software engineering text for emotion classification.

A recent study by Li et al.~\cite{li2023two} introduced the use of BERT~\cite{bert} as part of a two-stage framework for ambiguity detection, surpassing the performance of SEntiMoji in a GitHub emotion dataset curated by Imran et al.~\cite{imran2022data}. Another recent study by Bleyl et al.~\cite{bleyl2022emotion} found that BERT outperforms EMTk on Stack Overflow dataset curated by Novielli et al.~\cite{novielli2018gold}. This shift towards leveraging PTMs prompts a critical inquiry into the overall efficacy of such models in the nuanced task of emotion classification within software engineering contexts. Notably, Pre-trained Transformer Language Models like BERT~\cite{bert} and RoBERTa~\cite{roberta} have already demonstrated state-of-the-art results across various software engineering domains, including code completion, code review, bug localization, sentiment analysis, and toxicity detection~\cite{ciborowska2022fast, ciniselli2021empirical, linsentiment, sghaier2023multi, zhang2020sentiment, sarker2020benchmark, cassee2022sentiment}.

In light of these developments, this paper embarks on a comprehensive exploration of Pre-trained Transformers Language Models in emotion classification for software engineering texts. Through rigorous evaluation and comparison, we aim to shed light on the strengths, limitations, and potential implications of leveraging PTMs in this nuanced domain.

Our study utilizes two distinct datasets sourced from GitHub by Imran et al.~\cite{imran2022data} and Stack Overflow by Novielli et al.~\cite{novielli2018gold}, encompassing a diverse range of emotions expressed in software engineering contexts. We assess the effectiveness of six state-of-the-art PTMs, including four general domain models (BERT, RoBERTa, ALBERT and DeBERTa) and two specifically tailored for software engineering (CodeBERT and GraphCodeBERT). This evaluation is conducted against the benchmark set by the SEntiMoji model. EMTk and ESEM-E are excluded from this comparative analysis, aligning with findings reported by Chen et al.~\cite{sentimoji} indicating the superior performance of SEntiMoji.

Our investigation is guided by two primary research questions, each addressing specific facets of PTMs for emotion classification:

\smallskip
\noindent
{\sf
\textit{RQ1: How accurately can Pre-Trained Language Models classify emotions in software engineering texts compared to state-of-the-art model?}
}

\noindent
To address this question, we evaluate the performance of the six aforementioned state-of-the-art PTMs against the SEntiMoji model. Utilizing two distinct datasets, the GitHub dataset by Imran et al.\cite{imran2022data} and the Stack Overflow dataset by Novielli et al.\cite{novielli2018gold}, both annotated using the comprehensive model by Shaver et al.~\cite{Shaver}, we scrutinize the models' efficacy in capturing the nuances of emotions expressed in software engineering contexts. Our results suggest that the PTMs generally outperform state-of-the-art, average f1-score improvement ranges from 1.17\% to 16.79\%.

\smallskip
\noindent
{\sf
\textit{RQ2: Can integrating polarity features in the training improve Pre-Trained Language Models' ability for emotion classification?}
}

\noindent
Building upon observations by Chen et al.\cite{sentimoji} and Imran et al.\cite{imran2022data} regarding the impact of sentiment polarity on emotion classification errors, we focus on the potential benefits of incorporating polarity features during transformer model training. Specifically, we examine the extent to which positive and negative polarity features enhance the contextual understanding of PTMs for emotion classification in software engineering texts. We incorporate polarity features in attention layer of the PTMs. Our results suggest that they, indeed, improve the baseline PTMs further 1.0\% to 10.23\% in average f1-score metric.

The main contributions of this work are as follows:

\begin{itemize}
 \item To our best knowledge, this is the first study which leverages various PTMs for a large-scale comparative study against existing state-of-the-art tool for emotion classification in software engineering text.
 \item This is the first study which leverages polarity features to enhance PTMs for emotion classification in software engineering text.
\end{itemize}

We release the source code at URL: \url{https://anonymous.4open.science/status/SE_Emotion_PTM-3589}.
 \section{EXPERIMENT SETUP}

\subsection{State-of-the-art Models}

\noindent
\textbf{SEntiMoji}: SEntiMoji is a deep learning-based model and pretrained on GitHub data. The model is proposed by Chen et al.~\cite{sentimoji} and built on top of the DeepMoji~\cite{deepmoji} model. This flexible model can identify different emotion categorization schemes as well as sentiment classification.

\subsection{Compared PTMs} We compare 6 encoder-based PTMs: BERT, RoBERTa, DeBERTa, ALBERT, CodeBERT and GraphCodeBERT. Previous research shows that BERT, RoBERTa and ALBERT work well in SE affect analysis~\cite{zhang2020sentiment, batra2021bert}, while DeBERTa shows promising results in sentiment analysis~\cite{wang2022auto}. In addition to these four models, we compare against 2 domain-specific models that are widely used and proposed by Microsoft: CodeBERT and GraphCodeBERT -- which showed promising results in various SE tasks~\cite{zhou2021assessing, karmakar2021pre}. The models are explained below.

\noindent
\textbf{BERT~\cite{bert}:} Introduced by Google in 2018, BERT, pre-trained on a diverse corpus, including Wikipedia and Book Corpus, revolutionized natural language processing NLP with its bidirectional transformer architecture. This design allows BERT to capture context from both left and right contexts of a given word, making it highly effective for a diverse range of NLP tasks.

\noindent
\textbf{RoBERTa~\cite{roberta}:}
Developed by Meta AI, RoBERTa represents a refinement in transformer-based models. By modifying key hyperparameters and leveraging an extensive training dataset, RoBERTa enhances its performance on benchmark NLP tasks, showcasing improved language representation understanding.

\noindent
\textbf{DeBERTa~\cite{deberta}:}
An evolution of BERT and developed by Microsoft Research, DeBERTa focuses on enhancing the decoding process in language understanding tasks. It incorporates directional masks during training to capture bidirectional dependencies effectively, showcasing prowess in tasks requiring sequential reasoning.

\noindent
\textbf{ALBERT~\cite{albert}:}
Designed by Google Research, ALBERT introduces efficiency improvements to the BERT architecture without compromising representational power. This is achieved through parameter reduction techniques during training, showcasing resource-efficient yet highly effective language representations.

\noindent
\textbf{CodeBERT~\cite{codebert}:}
Tailored for programming languages and code understanding, CodeBERT, developed by Microsoft Research, is pre-trained on a large-scale dataset of programming tasks. This specialization allows CodeBERT to excel in source code-related tasks such as code summarization and variable naming.

\noindent
\textbf{GraphCodeBERT~\cite{graphcodebert}:} Developed by Microsoft Research, GraphCodeBERT is a transformer-based model designed for comprehending programming languages and code. Pre-trained on a vast dataset of programming tasks, it effectively captures the complex structures and semantics of code, serving as a proficient solution for source code-related tasks.

We use the publicly available versions of each model in Hugging Face~\cite{huggingface}. The model versions are shown in Table~\ref{tab:model_version}.

\begin{table}
    \small
    \centering
    \caption{Used model versions on Hugging Face}
    \begin{tabular}{|c|c|}
        \hline
        \textbf{Model} & \textbf{Version} \\
        \hline
        BERT & bert-base-uncased \\
        \hline
        RoBERTa & roberta-base \\
        \hline
        ALBERT & albert-base-v2 \\
        \hline
        DeBERTa & microsoft/deberta-v3-base \\
        \hline
        CodebERT & microsoft/codebert-base \\
        \hline
        GraphCodeBERT & microsoft/graphcodebert-base \\
        \hline
    \end{tabular}
    \label{tab:model_version}
\end{table}

\subsection{Datasets} We conduct experiment using two existing SE-emotion datasets. Both datasets are annotated using Shaver's emotion model~\cite{Shaver}. Shaver's emotion model has six basic emotion categories: Anger, Love, Fear, Joy, Sadness, and Surprise. The datasets are:

\noindent
\textbf{GitHub Emotion Dataset.} Imran et al.~\cite{imran2022data} curated a multi-label emotion dataset that is crawled from GitHub. The dataset consists of 2000 data points. The dataset contains 340 (17.0\%) Anger, 220 (11.0\%) Love, 198 (9.9\%) Fear, 422 (21.1\%) Joy, 274 (13.7\%) Sadness, and 328 (16.4\%) Surprise comments. The rest are neutral comments.

\noindent
\textbf{Stack Overflow Emotion Dataset.} Noveilli et al.~\cite{novielli2018gold} curated a multi-label emotion dataset from Stack Overflow Discussion. The dataset consists of 4800 data points. The dataset contains 882 (18\%) Anger, 1220 (25\%) Love, 106 (2\%) Fear, 491 (10\%) Joy, 230 (5\%) Sadness, and 45 (1\%) Surprise. The rest are neutral comments.

\noindent
For both datasets, we do simple text-preprocessing by lowercasing and removing null characters.

\begin{table*}[tb]
\centering
\small
\caption{Evaluation of PTMs on the Emotions Dataset (F1-score).}
\begin{tabular} {|l|l|c|c|c|c|c|c|c|c|}
\hline
    Dataset & Model & Anger & Love & Fear & Joy & Sadness & Surprise & Micro Avg. & Macro Avg.
    \\ \hline\hline
    & SEntiMoji & 0.460 & 0.642 & 0.377 & 0.556 & 0.629 & 0.458 & 0.530 & 0.521 \\ \cline{2-10}

    & BERT & 0.426 & 0.731 & 0.545 & 0.597 & 0.609 & \bf 0.639 & 0.585 & 0.591 \\
    &  \scriptsize (+/-) &  \scriptsize -7.49\% &  \scriptsize +13.83\% &  \scriptsize +44.55\%	&  \scriptsize +7.29\% &  \scriptsize -3.26\% &  \scriptsize +39.49\%  &  \scriptsize +10.37\% &  \scriptsize +13.56\% \\ \cline{2-10}
    & RoBERTa & 0.517  & \bf 0.774  & \bf 0.561  & 0.521  & \bf 0.653  & 0.511  & 0.575  & 0.590 \\
    &  \scriptsize (+/-) &  \scriptsize +12.44\% &  \scriptsize +20.60\% &  \scriptsize +48.66\% &  \scriptsize -6.37\% &  \scriptsize +3.85\% &  \scriptsize +11.55\% &  \scriptsize +8.45\% &  \scriptsize +13.28\%\\ \cline{2-10}
    & ALBERT & 0.446 & 0.753 & 0.357 & 0.447 & 0.631 & 0.602 & 0.538 & 0.539\\
    GitHub &  \scriptsize (+/-) &  \scriptsize-2.98\% &  \scriptsize+17.29\% &  \scriptsize-5.36\% & 	 \scriptsize-19.61\% &  \scriptsize+0.23\% &  \scriptsize+31.30\% &  \scriptsize+1.52\% &  \scriptsize+3.61\% \\ \cline{2-10}
    & DeBERTa & \bf 0.578 & 0.736 & 0.476 & \bf 0.605 & 0.642 & 0.611 &  \bf 0.610 &  \bf 0.608\\
    &  \scriptsize(+/-) &  \scriptsize+25.72\% &  \scriptsize+14.59\% &  \scriptsize+26.19\% &  \scriptsize+8.69\% &  \scriptsize+1.95\% &  \scriptsize+33.33\% &  \scriptsize+15.07\% &  \scriptsize+16.79\% \\
    \cline{2-10}
    & CodeBERT & 0.446 & 0.653 & 0.548 & 0.518 & 0.591 & 0.574 & 0.545 & 0.555\\
    &  \scriptsize(+/-) &  \scriptsize-3.01\% &  \scriptsize+1.73\% &  \scriptsize+45.21\% &  \scriptsize-6.86\% &  \scriptsize-6.09\% &  \scriptsize+25.19\% &  \scriptsize+2.95\% &  \scriptsize+6.62\% \\ \cline{2-10}
    & GraphCodeBERT  & 0.476 & 0.632 & 0.507 & 0.552 & 0.551 & 0.578 & 0.549 & 0.549\\
    &  \scriptsize(+/-) &  \scriptsize+3.52\% &  \scriptsize-1.62\% &  \scriptsize+34.27\% &  \scriptsize-0.74\% &  \scriptsize-12.43\% &  \scriptsize+26.14\% &  \scriptsize+3.60\% &  \scriptsize+5.53\% \\

    \cline{1-10}

    & SEntiMoji & 0.759 & 0.819 & 0.429 & 0.435 & 0.556 & 0.182 & 0.714 & 0.530 \\ \cline{2-10}

    & BERT & 0.769 & 0.851 & 0.545 & 0.597 & \bf 0.600 & 0.167 & 0.754 & 0.588 \\
    &  \scriptsize(+/-) &  \scriptsize+1.40\% &  \scriptsize+3.95\% &  \scriptsize+27.27\% &  \scriptsize+37.05\% &  \scriptsize+8.00\% &  \scriptsize-8.33\% &  \scriptsize+5.62\% &  \scriptsize+11.02\% \\ \cline{2-10}
    & RoBERTa & 0.786 & \bf 0.872 & 0.581 & 0.591 & \bf 0.600 & 0.1667 & \bf 0.758 & 0.599 \\
    &  \scriptsize(+/-) &  \scriptsize+3.60\% &  \scriptsize+6.49\% &  \scriptsize+35.48\% &  \scriptsize+35.77\% &  \scriptsize+8.00\% &  \scriptsize-8.33\% &  \scriptsize+6.23\% &  \scriptsize+13.13\% \\ \cline{2-10}
    & ALBERT & 0.762 & 0.845 & 0.579 & \bf 0.640 & 	0.545 & 0.133 & 0.747 & 0.584 \\
    Stack Overflow &  \scriptsize(+/-) &  \scriptsize+0.40\% &  \scriptsize+3.13\% &  \scriptsize+35.09\% &  \scriptsize+47.00\% &  \scriptsize-1.82\% &  \scriptsize-26.67\% &  \scriptsize+4.65\% &  \scriptsize+10.23\% \\ \cline{2-10}
    & DeBERTa & \bf 0.777 & 0.860 & \bf 0.591 & 0.604 & 0.598 & \bf 0.211 & 0.756 & \bf 0.607 \\
    &  \scriptsize(+/-) &  \scriptsize+2.49\% &  \scriptsize+5.04\% &  \scriptsize+37.88\% &  \scriptsize+38.68\% &  \scriptsize+7.59\% &  \scriptsize+15.79\% &  \scriptsize+5.90\% &  \scriptsize+14.52\% \\ \cline{2-10}
    & CodeBERT & 0.772 & 0.854 & 0.556 & 0.519 & 0.537 & 0.167 & 0.728 & 0.567 \\
    &  \scriptsize(+/-) &  \scriptsize+1.79\% &  \scriptsize+4.29\% &  \scriptsize+29.63\% &  \scriptsize+19.17\% &  \scriptsize-3.41\% &  \scriptsize-8.33\% &  \scriptsize+2.04\% &  \scriptsize+7.08\% \\ \cline{2-10}
    & GraphCodeBERT & 0.727	 & 0.836 & 0.514 & 0.559 & 0.521 & 0.154 & 0.722 & 0.552\\
    &  \scriptsize(+/-) &  \scriptsize-4.13\% &  \scriptsize+2.06\% &  \scriptsize+20.00\% &  \scriptsize+28.32\% &  \scriptsize-6.30\% &  \scriptsize-15.38\% &  \scriptsize+1.17\% &  \scriptsize+4.14\% \\
    \cline{2-10}

    \hline

\end{tabular}
\label{tab:classification_results_emotions}
\end{table*}

\subsection{Metrics}

We evaluate using F1-score -- which is commmonly used for this evaluation~\cite{sentimoji, imran2022data}. F1-score is the harmonic mean of Precision and Recall: $\text{\em F1-score} = 2 * \frac{Precision * Recall}{Precision + Recall}$. Precision measures the ratio of true positives to all positive predictions, while Recall calculates the ratio of true positives to all actual positives. Additionally, micro-averaged F1 considers the overall performance by aggregating individual class scores, treating each instance equally, whereas macro-averaged F1 computes the average across emotion classes, providing equal weight to each class regardless of their size.

\subsection{Classification Setup}

As both datasets are multi-label, we employ one-vs-all settings for all models. A one-vs.-all solution consists of N (here, N = 6) separate binary classifiers, each answering a separate classification question during training~\cite{one-vs-all}.

We divide the datasets into stratified train (80\%) and test (20\%) splits based on emotions using random sampling~\cite{botev2017variance}. For all 6 models, we will use the same train and test sets.

\section{EXPERIMENTS}

\subsection{RQ1: How accurately can Pre-Trained Language Models classify emotions in software engineering texts compared to state-of-the-art model?}

The results for emotion analysis on GitHub and Stack Overflow datasets, utilizing the PTMs are presented in Table~\ref{tab:classification_results_emotions}. The table includes F1-scores for individual emotion classes, micro and macro-averaged F1-scores, and the performance improvement over the baseline SEntiMoji model. From the table, it is evident that all models demonstrate performance improvement for both datasets in most cases.

\noindent
\textbf{GitHub Dataset}: DeBERTa attains the highest micro and macro-averaged F1-scores, exhibiting improvements of 15.07\% and 16.79\%, respectively. ALBERT performed worse than the models in both metrics. The two SE-specific models perform similar. Across all individual emotions, PTMs display improvement in 24 out of 36 instances. Surprise sees improvement for all six models, Love and Fear for five models, Anger and Sadness for three models, and Joy for two models. Breaking it down by individual models, BERT improves four emotions, RoBERTa five emotions, ALBERT three emotions, DeBERTa six emotions, CodeBERT three emotions, and GraphCodeBERT three emotions. Notably, DeBERTa demonstrates the most significant improvement in Anger (25.72\%), RoBERTa in Love (20.60\%), RoBERTa in Fear (48.66\%), DeBERTa in Joy (8.69\%), RoBERTa in Sadness (3.85\%), and BERT in Surprise (39.49\%).

\noindent
\textbf{Stack Overflow Dataset}: RoBERTa achieves the highest micro-averaged F1-score, while DeBERTa achieves the highest macro-averaged F1-score, improving by 6.23\% and 14.52\%, respectively. Between the two SE-specific models, CodeBERT outperforms GraphCodeBERT in both micro and macro-averaged F1-score. Across all individual emotions, PTMs show improvement in 27 out of 36 instances. Love, Fear, and Joy improve for all six models, Anger for five models, Sadness for three models, and Surprise for one model. It's important to note that the Stack Overflow dataset is more imbalanced than the GitHub dataset, with Surprise distribution being only 1\%, exhibiting the least improvement in all models.  Breaking it down by individual models, BERT improves five emotions, RoBERTa five emotions, ALBERT four emotions, DeBERTa six emotions, CodeBERT four emotions, and GraphCodeBERT three emotions. DeBERTa demonstrates the most significant improvement in Anger (3.60\%), RoBERTa in Love (6.49\%), DeBERTa in Fear (37.88\%), ALBERT in Joy (47.0\%), BERT and RoBERTa in Sadness (8.0\%), and DeBERTa in Surprise (15.79\%).

\noindent
\textbf{Error Analysis}: To identify model mistakes, we conduct an error analysis. Due to space constraints, we focus exclusively on the GitHub benchmark which has gone prior error analysis~\cite{imran2022data}. To understand where the PTMs commonly make mistake, we examine 67 cases where all six models agreed, yet the ground truth differed (9 false positives, 58 false negatives), distributed across emotions. For assessment, we employ Novielli et al.'s~\cite{novielli2018benchmark} error categorization, using a thematic approach. An author initially mapped errors, and a senior undergraduate student reviewed and resolved disagreements through discussion.

Echoing Imran et al.'s observations~\cite{imran2022data}, the prevalent categories are `General Error' and `Implicit Sentiment Polarity.' General errors, occurring 29/67 times, manifest when models misinterpret or struggle to comprehend lexical cues conveying emotions. For instance, the text ``\textit{Nice, this is more slick  {thumbs-up}}" is annotated as Joy. Another example, ``\textit{i'm actually surprised this didn't get flagged by the analyzer...}," annotated as Surprise, remains mispredicted by all models. The majority of Surprise (10/15), Joy (8/14), and Love (4/5) errors fall into the general errors category.

Implicit sentiment polarity errors occur 18/67 times. An example - ``\textit{Patiently waiting for any updates. [...]}" - is annotated as Sadness, with none of the models predicting it correctly. This category is particularly noticeable in Joy (6/14) and Sadness (5/11).

The third major error category is `Figurative Language' (9/67), which occurs when users use humor, irony, sarcasm, metaphors, etc to convey emotion. This category is noticeable among negative emotions (Anger 5/13 and Fear 3/9). For example, the following utterance ``\textit{[...] I understand what you mean by ``takeover" however it doesn't hurt to be a little more explicit.}'' - annotated as Anger but none of the models were able to detect it correctly.

Of note, in 13/67 cases, the utterances contain emojis which contribute in expressing emotions. The models possibly fail to capture them. For instance, ``\textit{And yes, there should be tests  {face-screaming-in-fear}  {face-screaming-in-fear}  {face-screaming-in-fear}}" - annotated as Fear, none of the models predict it accurately.

\smallskip
{\sf \noindent
\textbf{RQ1 Takeaway}: \color{blue} The transformer-based PTMs consistently outperform the state-of-the-art SEntiMoji model across all emotions for both datasets. DeBERTa and RoBERTa emerge as top performers. General PTMs outshine domain-specific ones in this context.
}

\subsection{RQ2: Can integrating polarity features in the training improve Pre-Trained Language Models' ability for emotion classification?}

\begin{table*}[tb]
\centering
\small
\caption{Evaluation of Polarity-enhanced PTMs on the Emotions Dataset (F1-score).}
\begin{tabular} {|l|l|c|c|c|c|c|c|c|c|}
\hline
    Dataset & Model & Anger & Love & Fear & Joy & Sadness & Surprise & Micro Avg. & Macro Avg.
    \\ \hline\hline

    \multirow{18}{0 pt}{GitHub} & BERT+Polarity & 0.484 & 0.733 & \bf 0.583 & 0.629 & 0.636 & \bf 0.661 & 619 & \bf 0.621 \\
    &  \scriptsize SEntiMoji (+/-) &  \scriptsize +5.19\% &  \scriptsize +14.13\% &  \scriptsize +54.58\%	&  \scriptsize +13.11\% &  \scriptsize +1.00\% &  \scriptsize +44.19\%  &  \scriptsize +16.98\% &  \scriptsize +19.28\% \\
    &  \scriptsize BERT (+/-) &  \scriptsize +13.71\% &  \scriptsize +0.26\% &  \scriptsize +6.94\%	&  \scriptsize +5.42\% &  \scriptsize +4.41\% &  \scriptsize +3.37\%  &  \scriptsize +5.99\% &  \scriptsize +5.04\% \\ \cline{2-10}
    & RoBERTa+Polarity & 0.475 & \bf 0.787 & 0.538 & 0.583 & 0.654 & 0.598 & 0.603 & 0.606 \\
    &  \scriptsize SEntiMoji (+/-) &  \scriptsize +3.31\% &  \scriptsize +22.63\% &  \scriptsize +42.47\%	&  \scriptsize +4.76\% &  \scriptsize +3.91\% &  \scriptsize +30.57\%  &  \scriptsize +13.81\% &  \scriptsize +16.39\% \\
    &  \scriptsize RoBERTa (+/-) &  \scriptsize -8.12\% &  \scriptsize +1.68\% &  \scriptsize -4.16\%	&  \scriptsize +11.89\% &  \scriptsize +0.06\% & \bf  \scriptsize +17.04\%  &  \scriptsize +4.94\% &  \scriptsize +2.75\% \\ \cline{2-10}
    & ALBERT+Polarity & 0.471 & 0.744 & 0.448 & 0.587 & \bf 0.674 & 0.561 & 0.580 & 0.581\\
    &  \scriptsize SEntiMoji (+/-) &  \scriptsize+2.30\% &  \scriptsize+15.92\% &  \scriptsize+18.66\% &  \scriptsize+5.46\% &  \scriptsize+7.07\% &  \scriptsize+22.34\% &  \scriptsize+9.49\% &  \scriptsize+11.54\% \\
    &  \scriptsize ALBERT (+/-) &  \scriptsize+5.45\% &  \scriptsize-1.16\% & \bf  \scriptsize+25.37\% & \bf  \scriptsize+31.19\% &  \scriptsize+6.83\% &  \scriptsize-6.82\% &  \scriptsize+7.86\% &  \scriptsize+7.65\% \\ \cline{2-10}
    & DeBERTa+Polarity  & \bf 0.588 & 0.680 & 0.507 & \bf 0.633 & 0.654 & 0.623 & \bf 0.620 & 0.614\\
    &  \scriptsize SEntiMoji (+/-) &  \scriptsize+27.88\% &  \scriptsize+5.99\% &  \scriptsize+34.37\% &  \scriptsize+13.85\% &  \scriptsize+3.91\% &  \scriptsize+35.85\% &  \scriptsize+17.09 &  \scriptsize+18.01 \\
    &  \scriptsize DeBERTa (+/-) &  \scriptsize+1.72\% &  \scriptsize-7.51\% &  \scriptsize+6.48\% &  \scriptsize+4.74\% &  \scriptsize+1.92\% &  \scriptsize+1.89\% &  \scriptsize+1.75\% &  \scriptsize+1.04\% \\ \cline{2-10}
    & CodeBERT+Polarity  & 0.565 & 0.691 & 0.576 & 0.530 & 0.607 & 0.640 & 0.595 & 0.601\\
    &  \scriptsize SEntiMoji (+/-) &  \scriptsize+22.80\% &  \scriptsize+7.62\% &  \scriptsize+52.58\% &  \scriptsize-4.66\% &  \scriptsize-3.57\% &  \scriptsize+39.64\% &  \scriptsize+12.38\% &  \scriptsize+15.55\% \\
    &  \scriptsize CodeBERT (+/-) & \bf  \scriptsize+26.61\% & \bf  \scriptsize+5.80\% &  \scriptsize+5.08\% &  \scriptsize+2.36\% &  \scriptsize+2.68\% &  \scriptsize+11.54\% & \bf  \scriptsize+9.16\% & \bf  \scriptsize+8.37\% \\ \cline{2-10}
    & GraphCodeBERT+Polarity & 0.514 & 0.654 & 0.551 & 0.570 & 0.598 & 0.521 & 0.563 & 0.568\\
    &  \scriptsize SEntiMoji (+/-) &  \scriptsize+11.63\% &  \scriptsize+1.91\% &  \scriptsize+45.94\% &  \scriptsize+2.42\% &  \scriptsize-5.01\% &  \scriptsize+13.70\% &  \scriptsize+6.22\% &  \scriptsize+9.09\% \\
    &  \scriptsize GraphCodeBERT (+/-) &  \scriptsize+7.84\% &  \scriptsize+3.58\% &  \scriptsize+8.70\% &  \scriptsize+3.19\% & \bf  \scriptsize+8.47\% &  \scriptsize-9.86\% &  \scriptsize+2.52\% &  \scriptsize+3.38\% \\

    \cline{1-10}

    \multirow{18}{0 pt}{Stack Overflow}  & BERT+Polarity & \bf 0.785 & 0.855 & 0.611 & 0.601 & 0.590 & 0.200 & 0.762 & 0.607\\
    &  \scriptsize SEntiMoji (+/-) &  \scriptsize+3.44\% &  \scriptsize+4.36\% &  \scriptsize+42.59\% &  \scriptsize+38.05\% &  \scriptsize+6.15\% &  \scriptsize+10.0\% &  \scriptsize+6.68\% &  \scriptsize+14.55\%\\
    &  \scriptsize BERT (+/-) &  \scriptsize+2.02\% &  \scriptsize+0.40\% &  \scriptsize+12.04\% &  \scriptsize+0.73\% &  \scriptsize-1.71\% &  \scriptsize+20.0\% &  \scriptsize+1.0\% &  \scriptsize+3.17\% \\ \cline{2-10}
    & RoBERTa+Polarity & 0.777 & \bf 0.880 & 0.650 & 0.594 & 0.575 & \bf 0.400 & \bf 0.767 & \bf 0.646 \\
    &  \scriptsize SEntiMoji (+/-) &  \scriptsize+2.45\% &  \scriptsize+7.47\% &  \scriptsize+51.67\% &  \scriptsize+36.45\% &  \scriptsize+3.45\% &  \scriptsize+120.0\% &  \scriptsize+7.51\% &  \scriptsize+21.93\% \\
    &  \scriptsize RoBERTa (+/-) &  \scriptsize-1.11\% &  \scriptsize+0.92\% &  \scriptsize+11.94\% &  \scriptsize+0.50\% &  \scriptsize-4.21\% &  \scriptsize+140.0\% &  \scriptsize+1.20\% &  \scriptsize+7.78\% \\ \cline{2-10}
    & ALBERT  & 0.777 & 0.844 & \bf 0.667 & 0.598 & \bf 0.644 & 0.167 & 0.757 & 0.616\\
    &  \scriptsize SEntiMoji (+/-) &  \scriptsize+2.44\% &  \scriptsize+3.09\% &  \scriptsize+55.56\% &  \scriptsize+37.31\% &  \scriptsize+16.0\% &  \scriptsize-8.33\% &  \scriptsize+6.08\% &  \scriptsize+16.30\% \\
    &  \scriptsize ALBERT (+/-) &  \scriptsize+2.04\% &  \scriptsize-0.04\% &  \scriptsize+15.15\% &  \scriptsize-6.59\% &  \scriptsize+18.15\% &  \scriptsize+25.0\% &  \scriptsize+1.36\% & \bf   \scriptsize+10.23\% \\ \cline{2-10}
    & DeBERTa+Polarity & 0.776 & 0.862 & 0.619 & \bf 0.643 & 0.623 & 0.222 & 0.766 & 0.624\\
    &  \scriptsize SEntiMoji (+/-) &  \scriptsize+2.29\% &  \scriptsize+5.27\% &  \scriptsize+44.44\% &  \scriptsize+47.74\% &  \scriptsize+12.21\% &  \scriptsize+22.22\% &  \scriptsize+7.36\% &  \scriptsize+17.84\% \\
    &  \scriptsize DeBERTa (+/-) &  \scriptsize-0.20\% &  \scriptsize+0.22\% &  \scriptsize+4.76\% &  \scriptsize+6.53\% &  \scriptsize+4.30\% &  \scriptsize+5.56\% &  \scriptsize+1.37\% &  \scriptsize+2.89\% \\ \cline{2-10}
    & CodeBERT+Polarity & 0.766 & 0.866 & 0.550 & 0.536 & 0.565 & 0.235 & 0.742 & 0.586 \\
    &  \scriptsize SEntiMoji (+/-) &  \scriptsize+0.95\% &  \scriptsize+5.70\% &  \scriptsize+28.33\% &  \scriptsize+23.0\% &  \scriptsize+1.65\% &  \scriptsize+29.41\% &  \scriptsize+3.99\% &  \scriptsize+10.64\% \\
    &  \scriptsize CodeBERT (+/-) &  \scriptsize-0.83\% &  \scriptsize+1.35\% &  \scriptsize-1.0\% &  \scriptsize+3.22\% &  \scriptsize+5.42\% &  \scriptsize+41.18\% & \bf  \scriptsize+1.91\% &  \scriptsize+3.32\% \\ \cline{2-10}
    & GraphCodeBERT+Polarity & 0.727 & 0.848 & 0.524 & 0.583 & 0.565 & 0.167 & 0.732 & 0.569 \\
    &  \scriptsize SEntiMoji (+/-) &  \scriptsize-4.20\% &  \scriptsize+3.56\% &  \scriptsize+22.22\% &  \scriptsize+33.98\% &  \scriptsize+1.65\% &  \scriptsize-8.33\% &  \scriptsize+2.48\% &  \scriptsize+7.38\% \\
    &  \scriptsize GraphCodeBERT (+/-) &  \scriptsize-0.07\% &  \scriptsize+1.47\% &  \scriptsize+1.85\% &  \scriptsize+4.42\% &  \scriptsize+8.48\% &  \scriptsize+8.33\% &  \scriptsize+1.29\% &  \scriptsize+3.11\% \\

    \hline

\end{tabular}
\label{tab:classification_results_emotions_polarity}
\end{table*}
In open-domain research, leveraging word polarity has proven effective for sentiment analysis~\cite{tian2020skep, zhang2023empirical, ke2020sentilare}. We aim to explore its applicability to emotion classification within the SE domain.

\subsubsection{Procedure} To integrate polarity features, we enhance PTMs through token-level attention, focusing on tokens associated with polarity words.

Initially, we employ natural language processing techniques, utilizing the Natural Language Toolkit (nltk) and SentiWordNet to extract word polarity. This involves a series of steps such as tokenization, part-of-speech tagging, and identification of words with discernible sentiment, resulting in a concise list of polarity words for each utterance.

\begin{figure}[t]
\centering
\includegraphics[width=0.3\textwidth]{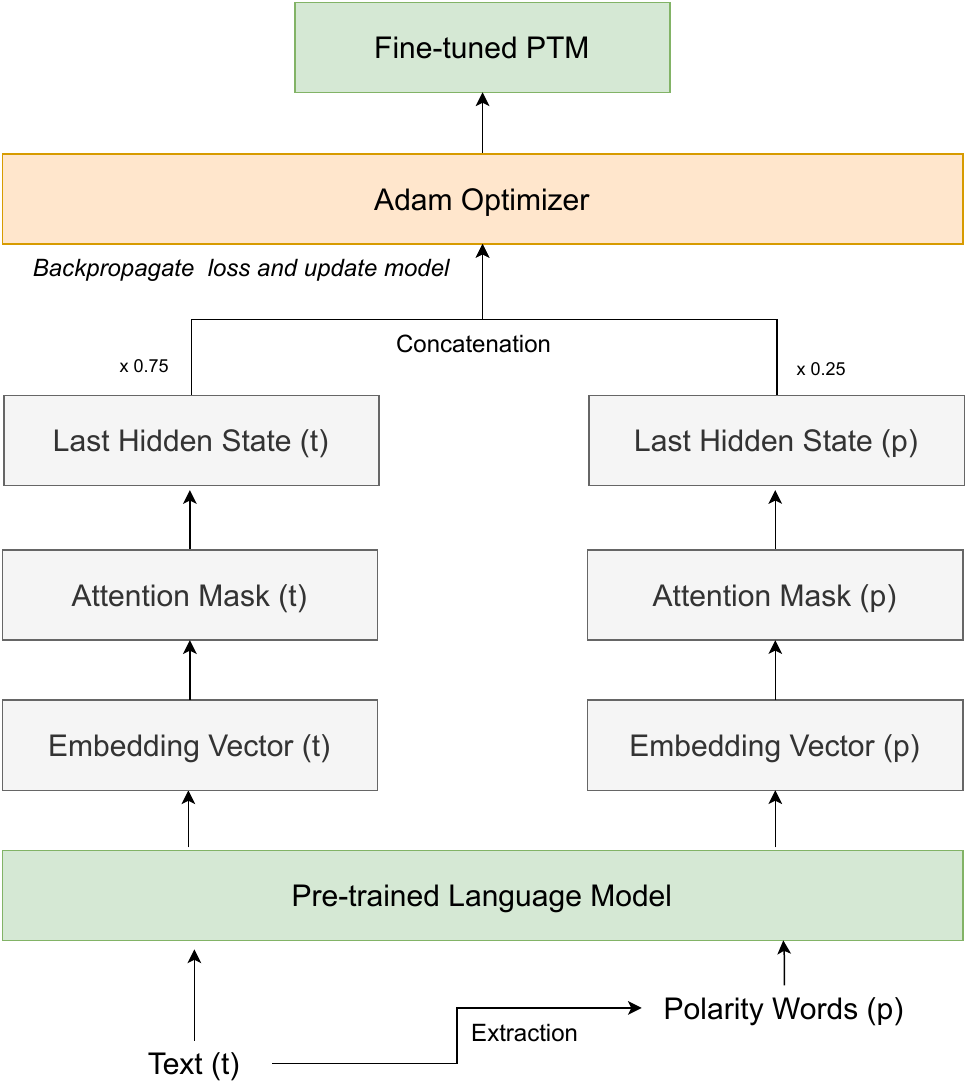}
\caption{Fine-tuning procedure using token-level attention adjustment of polarity words.}
\label{fig:finetuning}
\end{figure}

Subsequently, the model architecture is fine-tuned at the attention level, with a focus on polarity words. Attention weights are adjusted to assign greater significance to tokens linked with polarity words. This involves a strategic blending of attention weights related to the primary input text and those corresponding to polarity words. The adjustment ensures a heightened emphasis on the embeddings of polarity words, achieved by multiplying attention weights for primary input text's hidden states by 0.75 and those for polarity words' hidden states by 0.25, followed by concatenation. This modification enhances the model's sensitivity to sentiment-carrying terms during training, allowing for improved discernment of subtle variations in sentiment expression. Refer to Figure~\ref{fig:finetuning} for a visual representation of the token-level attention adjustment procedure.

\subsubsection{Results and Discussion} The results of emotion classification using polarity-enhanced PTMs are presented in Table~\ref{tab:classification_results_emotions_polarity}, along with the percentage difference in performance against a state-of-the-art model and baseline PTMs. The findings show that polarity-enhanced PTMs outperform the state-of-the-art SEntiMoji model and baseline PTMs.

\noindent
{\bf GitHub Dataset}: \textul{SEntiMoji baseline}: DeBERTa achieves the highest micro-averaged F1-score of 0.620, showing a 17.09\% improvement. On the macro-averaged F1-score metric, BERT achieves the highest score of 0.621, reflecting a 19.28\% improvement. Analyzing individual emotions, polarity-enhanced PTMs outperform SEntiMoji 33 out of 36 times, with improvements in Anger, Love, Fear, and Surprise for all six models, Joy for five models, and Sadness for four models. Anger achieves the highest F1-score of 0.588 by DeBERTa, marking a 27.88\% improvement from SEntiMoji, Love F1-score 0.787 by RoBERTa (+22.63\%), Fear F1-score 0.583 by BERT (+54.58\%), Joy F1-score 0.633 by BERT (+13.85\%), Sadness F1-score 0.674 by BERT (+7.07\%), and Surprise F1-score 0.661 by BERT (+44.19\%).

\textul{PTM baseline}:
The improvement on PTM baseline ranges from 2.52\% to 9.16\% on micro-averaged F1-score and from 1.04\% to 8.37\% on macro-averaged F1-score. The most improved PTM is ALBERT in both micro and macro-averaged F1-score. enhancement is observed consistently across individual emotions, with Joy and Sadness improving for all six models, Anger and Fear for five models, Love for four models, and Surprise for four models. Noteworthy improvements include a 26.61\% enhancement in Anger by CodeBERT, a 5.80\% improvement in Love by CodeBERT, a 25.37\% improvement in Fear by ALBERT, a 31.19\% improvement in Joy by ALBERT, an 8.47\% improvement in Sadness by GraphCodeBERT, and a 17.04\% improvement in Surprise by RoBERTa.

\noindent
{\bf Stack Overflow Dataset}: \textul{SEntiMoji baseline}: RoBERTa achieves the highest micro-averaged F1-score of 0.766, showing a 7.36\% improvement; as well as and the highest-averaged F1-score of 0.646, reflecting a 21.93\% improvement. Analyzing individual emotions, polarity-enhanced PTMs outperform SEntiMoji 33 out of 36 times, with improvements in Love, Fear, Joy, and Sadness for all six models, Anger for five models, and Surprise for four models. Anger achieved the highest F1-score of 0.777 by RoBERTa and ALBERT, marking a 2.44\% improvement from SEntiMoji, Love F1-score 0.880 by RoBERTa (+7.47\%), Fear F1-score 0.667 by ALBERT (+55.56\%), Joy F1-score 0.643 by DeBERTa (+47.74\%), Sadness F1-score 0.664 by BERT (+16.0\%), and Surprise F1-score 0.400 by BERT (+120.0\%).

\textul{PTM baseline}:
The improvement on PTM baseline ranges from 1.0\% to 1.91\% on micro-averaged F1-score and from 3.11\% to 10.23\% on macro-averaged F1-score.
The most improved PTM is ALBERT in macro-averaged F1-score metric and CodeBERT in macro-averaged F1-score metric.
While the improvement in Stack Overflow dataset against baseline PTMs is less compared to the GitHub dataset, this enhancement is observed consistently across individual emotions, with Surprise improving for all six models, Love, Fear and Joy for five models, Sadness for four models, and Anger for four models. Noteworthy improvements include a 2.04\% enhancement in Anger by ALBERT, a 1.47\% improvement in Love by GraphCodeBERT, a 15.15\% improvement in Fear by ALBERT, a 6.53\% improvement in Joy by DeBERTa, an 18.15\% improvement in Sadness by ALBERT, and a 140.0\% improvement in Surprise by RoBERTa.

The average PTM improvement in both dataset are similar to the general domain sentiment analysis results~\cite{tian2020skep, zhang2023empirical}.

{\bf Error Analysis}: Similar to RQ1, we look into GitHub dataset for this case as well. In RQ1 error analysis, we observe that in 67 cases all models predict incorrectly. Out of those 67 cases, we find that 40 cases still produce erroneous results. However, in rest 27 cases, at least one model predict correctly. The most prominent resolved error categories are: 13/29 (44.82\%) general errors, 9/18 (50\%) implicit sentiment polarity, and 2/3 politeness. For example, ``\textit{i'm actually surprised this didn't get flagged by the analyzer...}," - this utterance is now correctly predicted by CodeBERT.

The least improved error category is `Pragmatics'. Pragmatics is the error category when the classifiers deal with context information. That is often human annotators consider external facts for annotation. For example, consider the following utterance, \textit{``[...] This change makes it the same as the line above and I don't see any reason to have two lines that are showing the same thing.''} - is annotated as Anger as the commentator is annoyed/disagreed with the change in code. 6/7 (85.71\%) Pragmatics related errors remains unresolved. Another least improved error category is `Figurative Language' - 6/9 (66\%) utterances predictions remained unchanged. For example, this following utterance, \textit{``[...] We need to add this test coverage. It's just not `urgent'.  {neutral-face}''} - which expresses sarcasm and annotated as Anger. The models predict it incorrectly.

We observe that there still remains a considerable amount of utterances (10/13) that are misclassified contain emojis. For example, the previously mentioned utterance, ``\textit{And yes, there should be tests  {face-screaming-in-fear}  {face-screaming-in-fear}  {face-screaming-in-fear}}" - all models still predict incorrectly.

\smallskip
{\sf \noindent
\textbf{RQ2 Takeaway}: \color{blue}
% Polarity-enhanced PTMs consistently outperform both the state-of-the-art SEntiMoji model and baseline PTMs in both datasets. The improvements are evident in micro-averaged and macro-averaged F1-scores, with substantial improvements observed for individual emotions for different models. While the findings suggest that incorporating sentiment polarity information into PTMs can enhance their performance in recognizing emotions, they do not always help to capture emotions, espectively when it is context dependant.
Polarity-enhanced PTMs consistently outperform the SEntiMoji model and baseline PTMs in both datasets. Notable improvements are seen in micro- and macro-averaged F1-scores, along with substantial enhancements for individual metric across various models. While the findings indicate that integrating sentiment polarity improves PTM performance, it do not always help to capture context-dependent emotions.
}

\section{DISCUSSION}

In this section, based on our experiments, we summarize the key lessons learned and outline promising directions for future work related to emotion classification in SE.

\subsection{Lessons Learned}

Our study yields several key insights into the application of PTMs for emotion classification in SE texts. Firstly, general domain PTMs such as BERT, RoBERTa, and DeBERTa consistently outperform specialized models. This suggests that the pre-training objectives and textual domains carry greater relevance than task-specific tuning in the emotion classification task within SE text.

Secondly, the incorporation of polarity features during fine-tuning consistently enhances performance, emphasizing the value of sentiment awareness. However, challenges persist, especially with negative emotions like Anger and Fear, which pose difficulties in context understanding.

Thirdly, no single model excels across all emotions in all metrics, necessitating careful benchmarking and tradeoff analysis. Common error categories, such as implicit polarity, figurative language, and pragmatics, underscore the challenges in understanding contextual gaps in nuanced emotion recognition. Addressing these complexities requires further customization. For example, Imran et al.~\cite{imran2023shedding} addressed the issue of understanding figurative language in SE text using contrastive learning.

Lastly, persistent challenges arise in handling emojis, as models struggle to interpret emotive signals. Incorporating dedicated approaches to handle emojis may offer mitigation strategies.

% Our study provides several key lessons regarding the application of PTMs for emotion classification in SE texts:
%
% \begin{itemize}
% \item General domain PTMs like BERT, RoBERTa and DeBERTa consistently outperform specialized models like CodeBERT and GraphCodeBERT. This indicates that the pre-training objectives and textual domains hold greater relevance than task-specific tuning in emotion classification task in SE text.
%
% \item The inclusion of polarity features during PTM fine-tuning consistently improves performance, highlighting the value of sentiment awareness. However, certain emotions, particularly negative ones like Anger, Fear, and Sadness, still pose challenges in terms of context understanding, figurative language, and pragmatics.
%
% \item No single PTM excels across all metrics, necessitating careful benchmarking and tradeoff analysis.
%
%
% \item Common error categories include general errors, implicit polarity, figurative language, and contextual gaps, underscore the difficulties in nuanced emotion recognition, requiring further customization to address these limitations.
%
% \item Persistent challenges arise in handling emojis, with models struggling to interpret emotive signals. Dedicated approaches to handle emojis may mitigate this issue.
%
% \end{itemize}

\subsection{Future Directions}

Building on lessons learned, future work can focus on establishing more benchmark datasets, particularly emphasizing implicit and context-dependent emotions.

Investigating hierarchical emotion classification and joint modeling of related polarity dimensions merits investigation. Recognizing that emotion classification often treats each category separately, a hierarchical framework may enhance performance by first identifying broad emotional valence (positive, negative, or neutral) before classifying specific categories, as demonstrated by Li et al.~\cite{li2023two}.

In SE-specific models, exploring pre-training with polarity-word masking, emojis, and emoticons atop general word masking is warranted, given promising results in the general domain for sentiment analysis~\cite{zhou2020sentix}. Research also suggests that aspect-based sentiment analysis (ABSA)-enhanced PTMs outperform baseline PTMs~\cite{yang2023pyabsa, zhang2023empirical}. As polarity-enhanced PTMs have already shown superiority over baseline PTMs, exploring ABSA-enhanced PTMs may further improve classifiers.

Additionally, investigating the multi-modal fusion of text and emojis cues during the pre-training/fine-tuning step could enhance the interpretation of emotive expressions. Exploring generative language models like GPT-4 and LLaMA for emotion detection using zero-shot and few-shot learning presents an intriguing direction. Recent advancements in these models enable synthesizing natural language responses, facilitating data augmentation and prompting techniques for improved classification~\cite{bayer2022survey, kocon2023chatgpt, koptyra2023clarin, nedilko2023generative}. Focusing on detecting the most relevant emotions that may harm productivity, such as Frustration and Disappointment, holds merit~\cite{wrobel2013emotions}. For example, a recent case study delved into understanding Frustration in OSS text using generative models in zero-shot learning~\cite{imran2023uncovering}.

 \section{RELATED WORK}

On a large scale the related work can be divided into two parts: emotions in SE texts and PTMs in SE tasks.

\subsection{Emotions in Text-Based Software Engineering Communication}

Pletea et al. explored the emotions expressed by developers during security discussions on GitHub~\cite{pletea2014security}. Ortu et al. analyzed the emotional expressions in JIRA issue comments and their impact on collaboration and software development outcomes~\cite{ortu2016emotional}. Novielli et al. annotated a benchmark for emotion annotation in Stack Overflow using Shaver's~\cite{Shaver} emotion model~\cite{novielli2018gold}. Calefato et al. developed a toolkit for emotion recognition from text~\cite{emtk}. Mäntylä et al. explored the possibilities of using Valence, Arousal, and Dominance (VAD) dimensions~\cite{russell1977evidence} for detecting burnout and productivity~\cite{mantyla2016mining}. Islam et al. proposed MarValous -- a machine learning method for sensing emotions in the VAD space~\cite{islam2019marvalous}. Werder et al. proposed a method for extracting emotions from GitHub repositories~\cite{werder2018meme}. Destefanis et al. studied the measurement of affects in 370K GitHub issue comments to understand the emotional impact of commenters on the overall affect of GitHub issues~\cite{destefanis2018measuring}. Ortu et al. conducted a mining analysis on finding emotional state of contributors in GitHub repositories~\cite{ortu2018mining}.
Venigalla et al. explored the emotions in software documentation~\cite{venigalla2021understanding}.
Neupane et al. investigated emotion dynamics in software development \cite{neupane2019emod}.

Chen et al. utilized emojis for sentiment analysis and emotion detection in developers' communication through deep neural networks~\cite{sentimoji}. Rong et al. empirically studied emoji usage in software development communication, exploring frequency and diversity across different channels to understand developers' emotional expression~\cite{rong2022empirical}. Bleyl et al. applied a BERT-based emotion recognition model to Stack Overflow posts, aiming to identify developers' expressed emotions~\cite{bleyl2022emotion}. Imran et al. proposed an enhanced model based on Shaver's approach for software engineering text and introduced three data augmentation techniques to enhance Software Engineering emotion detection tool performance~\cite{imran2022data}. Tulli et al.~\cite{tulili2023burnout} did a literature review on burnout in software engineering.

\subsection{PTMs in Software Engineering Tasks}

The utilization of PTMs has gained significant attention in software engineering text, offering a novel approach to address various challenges and enhance the performance of diverse tasks. One notable area of research involves the application of PTMs in code-related tasks. Researchers have explored the effectiveness of models like in understanding and generating code snippets. These studies aim to improve code completion, summarization, comprehension and documentation~\cite{ciniselli2021empirical, gao2023code, haque2022semantic, macneil2023experiences, rai2022review}. Lanaguage models have shown promise in tasks related to software bugs and assessing code quality~\cite{ciborowska2022fast, ardimento2020using, wang2023clebpi, ali2023bert, zhou2021assessing, keim2020does}. There also has been study on understanding code attentions in BERT~\cite{sharma2022exploratory}. Researchers also conduct studies on benchmarking PTMs on tag recommendation in Stack Overflow text, sentiment analysis in SE text and software aspect-based API review text~\cite{zhang2020sentiment, yang2022aspect, he2022ptm4tag}.

In this study, we do comprehensive benchmark analysis of emotion classification task in SE text using PTMs and we also apply approaches in changing attention layer for further improving the performances.

 \section{THREATS TO VALIDY} Several limitations could affect the accurate interpretation of our results. We outline and detail each of these limitations below.

\noindent{\textul{Construct validity.}} Construct validity assesses how well the study measures the concepts and constructs it aims to evaluate. While potential issues may arise from manually annotating the error analysis in RQ1, we mitigated this by employing thematic analysis to address any possible problems.

\noindent{\textul{Internal validity.}} Internal validity focuses on the extent to which the study's findings can be attributed to the manipulation of the independent variable. Threats may arise from mistakes in our experiments, but we have provided a replication package and outputs for independent verifications. Another potential threat is the absence of cross-validation on smaller datasets, which we addressed by using stratified sampling for representativeness and an 80\%-20\% train-test split. There's also a potential threat related to data leakage in SEntiMoji models against the GitHub benchmark dataset, but this is unlikely as SEntiMoji was pre-trained before 2019~\cite{sentimoji}, and data points of the GitHub dataset came afterward~\cite{imran2022data}. However, verification of this is beyond the scope of our study given that not all pre-trained data is released by the authors~\cite{sentimoji}.

\noindent{\textul{External validity.}} External validity concerns the generalization of our study's findings to other settings and contexts.
Our results may not extend beyond the specific models, datasets, and domains of GitHub and Stack Overflow. However, we utilized diverse pre-trained models and the most comprehensive datasets available in this field. Further investigation is necessary to validate our results beyond the datasets and pre-trained models used in our study.

\section{CONCLUSION}
This paper presented a comprehensive comparative analysis of state-of-the-art Pre-trained Transformer Language Models for emotion classification in software engineering texts. Through evaluation on two datasets, sourced from GitHub and Stack Overflow and annotated using Shaver's emotion model, we address two key research questions. First, we examined the accuracy of PTMs in classifying emotions compared to SEntiMoji, we found consistent enhancements with models like BERT, DeBERTa and RoBERTa achieving upto 16.79\% improvements in terms of macro and micro-averaged F1 scores. The results showcase the prowess of general domain PTMs, with domain specific CodeBERT and GraphCodeBERT underperforming. Analysis of common errors revealed struggles in handling general comprehension, implicit polarity, figurative language and contextual pragmatics. Next, we incorporated polarity features during PTM finetuning and demonstrated additional gains over both SEntiMoji and baseline PTMs. This validates the benefit of augmenting with explicit polarity information for nuanced emotion classification. However, errors linked to figurative language and pragmatics persisted.
In conclusion, this study provides a robust benchmark of diverse PTMs in a complex affective analysis task within software engineering, showcasing their effectiveness not only in sentiment analysis but also in fine-grained emotion recognition. These findings underscore the potential of PTMs in advancing empathetic software intelligence, with opportunities for further refinement to address contextual gaps.

\bibliographystyle{IEEEtran}
\bibliography{references}

\begin{thebibliography}{10}
\providecommand{\url}[1]{#1}
\csname url@samestyle\endcsname
\providecommand{\newblock}{\relax}
\providecommand{\bibinfo}[2]{#2}
\providecommand{\BIBentrySTDinterwordspacing}{\spaceskip=0pt\relax}
\providecommand{\BIBentryALTinterwordstretchfactor}{4}
\providecommand{\BIBentryALTinterwordspacing}{\spaceskip=\fontdimen2\font plus
\BIBentryALTinterwordstretchfactor\fontdimen3\font minus
  \fontdimen4\font\relax}
\providecommand{\BIBforeignlanguage}[2]{{%
\expandafter\ifx\csname l@#1\endcsname\relax
\typeout{** WARNING: IEEEtran.bst: No hyphenation pattern has been}%
\typeout{** loaded for the language `#1'. Using the pattern for}%
\typeout{** the default language instead.}%
\else
\language=\csname l@#1\endcsname
\fi
#2}}
\providecommand{\BIBdecl}{\relax}
\BIBdecl

\bibitem{novielli2023emotion}
N.~Novielli and A.~Serebrenik, ``Emotion analysis in software ecosystems,'' in
  \emph{Software Ecosystems: Tooling and Analytics}.\hskip 1em plus 0.5em minus
  0.4em\relax Springer, 2023.

\bibitem{lin2022opinion}
B.~Lin, N.~Cassee, A.~Serebrenik, G.~Bavota, N.~Novielli, and M.~Lanza,
  ``Opinion mining for software development: a systematic literature review,''
  \emph{ACM TOSEM}, vol.~31, 2022.

\bibitem{esem-e}
A.~Murgia, M.~Ortu, P.~Tourani, B.~Adams, and S.~Demeyer, ``An exploratory
  qualitative and quantitative analysis of emotions in issue report comments of
  open source systems,'' \emph{Empirical Software Engineering}, 2018.

\bibitem{emtk}
F.~Calefato, F.~Lanubile, N.~Novielli, and L.~Quaranta, ``Emtk-the emotion
  mining toolkit,'' in \emph{2019 IEEE/ACM 4th International Workshop on
  SEmotion}.\hskip 1em plus 0.5em minus 0.4em\relax IEEE, 2019.

\bibitem{sentimoji}
Z.~Chen, Y.~Cao, H.~Yao, X.~Lu, X.~Peng, H.~Mei, and X.~Liu, ``Emoji-powered
  sentiment and emotion detection from software developers’ communication
  data,'' \emph{ACM TOSEM}, 2021.

\bibitem{ortu2016emotional}
M.~Ortu, A.~Murgia, G.~Destefanis, P.~Tourani, R.~Tonelli, M.~Marchesi, and
  B.~Adams, ``The emotional side of software developers in jira,'' in
  \emph{Proceedings of the 13th MSR}, 2016.

\bibitem{novielli2018gold}
N.~Novielli, F.~Calefato, and F.~Lanubile, ``A gold standard for emotion
  annotation in stack overflow,'' in \emph{2018 IEEE/ACM 15th MSR}.\hskip 1em
  plus 0.5em minus 0.4em\relax IEEE, 2018.

\bibitem{li2023two}
J.~Li, Y.~Lei, S.~Li, H.~Zhou, Y.~Yu, Z.~Jia, Y.~Ma, and T.~Wang, ``A two-stage
  framework for ambiguous classification in software engineering,'' in
  \emph{2023 IEEE 34th ISSRE}.\hskip 1em plus 0.5em minus 0.4em\relax IEEE,
  2023.

\bibitem{bert}
J.~Devlin, M.-W. Chang, K.~Lee, and K.~Toutanova, ``Bert: Pre-training of deep
  bidirectional transformers for language understanding,'' \emph{arXiv}, 2018.

\bibitem{imran2022data}
M.~M. Imran, Y.~Jain, P.~Chatterjee, and K.~Damevski, ``Data augmentation for
  improving emotion recognition in software engineering communication,'' in
  \emph{Proceedings of the 37th ASE}, 2022.

\bibitem{bleyl2022emotion}
D.~Bleyl and E.~K. Buxton, ``Emotion recognition on stackoverflow posts using
  bert,'' in \emph{2022 IEEE International Conference on Big Data}.\hskip 1em
  plus 0.5em minus 0.4em\relax IEEE, 2022.

\bibitem{roberta}
Y.~Liu, M.~Ott, N.~Goyal, J.~Du, M.~Joshi, D.~Chen, O.~Levy, M.~Lewis,
  L.~Zettlemoyer, and V.~Stoyanov, ``Roberta: A robustly optimized bert
  pretraining approach,'' \emph{arXiv}, 2019.

\bibitem{ciborowska2022fast}
A.~Ciborowska and K.~Damevski, ``Fast changeset-based bug localization with
  bert,'' in \emph{Proceedings of the 44th ICSE}, 2022.

\bibitem{ciniselli2021empirical}
M.~Ciniselli, N.~Cooper, L.~Pascarella, D.~Poshyvanyk, M.~D. Penta, and
  G.~Bavota, ``An empirical study on the usage of bert models for code
  completion,'' in \emph{2021 IEEE/ACM 18th MSR}.\hskip 1em plus 0.5em minus
  0.4em\relax IEEE, 2021.

\bibitem{linsentiment}
B.~Lin, F.~Zampetti, G.~Bavota, M.~Di~Penta, M.~Lanza, and R.~Oliveto,
  ``Sentiment analysis for software engineering: How far can we go?'' in
  \emph{Proceedings of the 40th ICSE}, 2018.

\bibitem{sghaier2023multi}
O.~B. Sghaier and H.~Sahraoui, ``A multi-step learning approach to assist code
  review,'' in \emph{2023 IEEE SANER}.\hskip 1em plus 0.5em minus 0.4em\relax
  IEEE, 2023.

\bibitem{zhang2020sentiment}
T.~Zhang, B.~Xu, F.~Thung, S.~A. Haryono, D.~Lo, and L.~Jiang, ``Sentiment
  analysis for software engineering: How far can pre-trained transformer models
  go?'' in \emph{2020 IEEE ICSME}.\hskip 1em plus 0.5em minus 0.4em\relax IEEE,
  2020.

\bibitem{sarker2020benchmark}
J.~Sarker, A.~K. Turzo, and A.~Bosu, ``A benchmark study of the contemporary
  toxicity detectors on software engineering interactions,'' in \emph{2020 27th
  APSEC}.\hskip 1em plus 0.5em minus 0.4em\relax IEEE, 2020.

\bibitem{cassee2022sentiment}
N.~Cassee, ``Sentiment in software engineering: detection and application,'' in
  \emph{Proceedings of the 30th ACM Joint European Software Engineering
  Conference and Symposium on the Foundations of Software Engineering}, 2022.

\bibitem{Shaver}
P.~Shaver, J.~Schwartz, D.~Kirson, and C.~O'connor, ``Emotion knowledge:
  further exploration of a prototype approach.'' \emph{Journal of personality
  and social psychology}, 1987.

\bibitem{deepmoji}
B.~Felbo, A.~Mislove, A.~S{\o}gaard, I.~Rahwan, and S.~Lehmann, ``Using
  millions of emoji occurrences to learn any-domain representations for
  detecting sentiment, emotion and sarcasm,'' in \emph{Proceedings of the 2017
  Conference on EMNLP}, 2017.

\bibitem{batra2021bert}
H.~Batra, N.~S. Punn, S.~K. Sonbhadra, and S.~Agarwal, ``Bert-based sentiment
  analysis: A software engineering perspective,'' in \emph{Database and Expert
  Systems Applications: 32nd International Conference, DEXA 2021}.\hskip 1em
  plus 0.5em minus 0.4em\relax Springer-Verlag, 2021.

\bibitem{wang2022auto}
T.~Wang, B.~Sun, and Y.~Tong, ``Auto-absa: automatic detection of aspects in
  aspect-based sentiment analysis,'' \emph{arXiv preprint arXiv:2202.00484},
  2022.

\bibitem{zhou2021assessing}
X.~Zhou, D.~Han, and D.~Lo, ``Assessing generalizability of codebert,'' in
  \emph{2021 IEEE ICSME}.\hskip 1em plus 0.5em minus 0.4em\relax IEEE, 2021.

\bibitem{karmakar2021pre}
A.~Karmakar and R.~Robbes, ``What do pre-trained code models know about code?''
  in \emph{2021 36th IEEE/ACM International Conference on Automated Software
  Engineering (ASE)}.\hskip 1em plus 0.5em minus 0.4em\relax IEEE, 2021, pp.
  1332--1336.

\bibitem{deberta}
P.~He, X.~Liu, J.~Gao, and W.~Chen, ``Deberta: Decoding-enhanced bert with
  disentangled attention,'' \emph{arXiv}, 2020.

\bibitem{albert}
Z.~Lan, M.~Chen, S.~Goodman, K.~Gimpel, P.~Sharma, and R.~Soricut, ``Albert: A
  lite bert for self-supervised learning of language representations,''
  \emph{arXiv}, 2019.

\bibitem{codebert}
Z.~{Feng}, D.~{Guo}, D.~{Tang}, N.~{Duan}, X.~{Feng}, M.~{Gong}, L.~{Shou},
  B.~{Qin}, T.~{Liu}, D.~{Jiang} \emph{et~al.}, ``Codebert: A pre-trained model
  for programming and natural languages,'' in \emph{EMNLP 2020}, 2020.

\bibitem{graphcodebert}
D.~Guo, S.~Ren, S.~Lu, Z.~Feng, D.~Tang, S.~Liu, L.~Zhou, N.~Duan,
  A.~Svyatkovskiy, S.~Fu \emph{et~al.}, ``Graphcodebert: Pre-training code
  representations with data flow,'' \emph{arXiv}, 2020.

\bibitem{huggingface}
\BIBentryALTinterwordspacing
(2023) Hugging face. [Online]. Available: \url{https://huggingface.co/}
\BIBentrySTDinterwordspacing

\bibitem{one-vs-all}
\BIBentryALTinterwordspacing
(2023) One vs all. [Online]. Available:
  \url{https://developers.google.com/machine-learning/crash-course/multi-class-neural-networks/one-vs-all}
\BIBentrySTDinterwordspacing

\bibitem{botev2017variance}
Z.~Botev and A.~Ridder, ``Variance reduction,'' \emph{Wiley statsRef:
  Statistics reference online}, 2017.

\bibitem{novielli2018benchmark}
N.~Novielli, D.~Girardi, and F.~Lanubile, ``A benchmark study on sentiment
  analysis for software engineering research,'' in \emph{Proceedings of the
  15th MSR}, 2018.

\bibitem{tian2020skep}
H.~Tian, C.~Gao, X.~Xiao, H.~Liu, B.~He, H.~Wu, H.~Wang, and F.~Wu, ``Skep:
  Sentiment knowledge enhanced pre-training for sentiment analysis,''
  \emph{arXiv}, 2020.

\bibitem{zhang2023empirical}
Y.~Zhang, Y.~Yang, B.~Liang, S.~Chen, B.~Qin, and R.~Xu, ``An empirical study
  of sentiment-enhanced pre-training for aspect-based sentiment analysis,'' in
  \emph{Findings of the ACL 2023}, 2023.

\bibitem{ke2020sentilare}
P.~Ke, H.~Ji, S.~Liu, X.~Zhu, and M.~Huang, ``Sentilare: Sentiment-aware
  language representation learning with linguistic knowledge,'' in
  \emph{Proceedings of the 2020 Conference on EMNLP}, 2020.

\bibitem{imran2023shedding}
M.~M. Imran, P.~Chatterjee, and K.~Damevski, ``Shedding light on software
  engineering-specific metaphors and idioms,'' \emph{arXiv preprint
  arXiv:2312.10297}, 2023.

\bibitem{zhou2020sentix}
J.~Zhou, J.~Tian, R.~Wang, Y.~Wu, W.~Xiao, and L.~He, ``Sentix: A
  sentiment-aware pre-trained model for cross-domain sentiment analysis,'' in
  \emph{Proceedings of the 28th international conference on computational
  linguistics}, 2020.

\bibitem{yang2023pyabsa}
H.~Yang, C.~Zhang, and K.~Li, ``Pyabsa: A modularized framework for
  reproducible aspect-based sentiment analysis,'' in \emph{Proceedings of the
  32nd ACM CIKM}, 2023.

\bibitem{bayer2022survey}
M.~Bayer, M.-A. Kaufhold, and C.~Reuter, ``A survey on data augmentation for
  text classification,'' \emph{ACM Computing Surveys}, 2022.

\bibitem{kocon2023chatgpt}
J.~Kocon, I.~Cichecki, O.~Kaszyca, M.~Kochanek, D.~Szyd{\l}o, J.~Baran,
  J.~Bielaniewicz, M.~Gruza, A.~Janz, K.~Kanclerz \emph{et~al.}, ``Chatgpt:
  Jack of all trades, master of none,'' \emph{Information Fusion}, 2023.

\bibitem{koptyra2023clarin}
B.~Koptyra, A.~Ngo, {\L}.~Radli{\'n}ski, and J.~Koco{\'n}, ``Clarin-emo:
  Training emotion recognition models using human annotation and chatgpt,'' in
  \emph{International Conference on Computational Science}.\hskip 1em plus
  0.5em minus 0.4em\relax Springer, 2023.

\bibitem{nedilko2023generative}
A.~Nedilko, ``Generative pretrained transformers for emotion detection in a
  code-switching setting,'' in \emph{Proceedings of the 13th Workshop on
  Computational Approaches to Subjectivity, Sentiment, \& Social Media
  Analysis}, 2023.

\bibitem{wrobel2013emotions}
M.~R. Wrobel, ``Emotions in the software development process,'' in \emph{2013
  6th International Conference on Human System Interactions (HSI)}.\hskip 1em
  plus 0.5em minus 0.4em\relax IEEE, 2013.

\bibitem{imran2023uncovering}
M.~M. Imran, P.~Chatterjee, and K.~Damevski, ``Uncovering the causes of
  emotions in software developer communication using zero-shot llms,''
  \emph{arXiv preprint arXiv:2312.09731}, 2023.

\bibitem{pletea2014security}
D.~Pletea, B.~Vasilescu, and A.~Serebrenik, ``Security and emotion: sentiment
  analysis of security discussions on github,'' in \emph{Proceedings of the
  11th working conference on mining software repositories}, 2014.

\bibitem{russell1977evidence}
J.~A. Russell and A.~Mehrabian, ``Evidence for a three-factor theory of
  emotions,'' \emph{Journal of research in Personality}, 1977.

\bibitem{mantyla2016mining}
M.~Mantyla, B.~Adams, G.~Destefanis, D.~Graziotin, and M.~Ortu, ``Mining
  valence, arousal, and dominance: possibilities for detecting burnout and
  productivity?'' in \emph{Proceedings of the 13th MSR}, 2016.

\bibitem{islam2019marvalous}
M.~R. {Islam}, M.~K. {Ahmmed}, and M.~F. {Zibran}, ``Marvalous: Machine
  learning based detection of emotions in the valence-arousal space in software
  engineering text,'' in \emph{Proceedings of the 34th ACM/SIGAPP SAC}, 2019.

\bibitem{werder2018meme}
K.~Werder and S.~Brinkkemper, ``Meme: toward a method for emotions extraction
  from github,'' in \emph{Proceedings of the 3rd International Workshop on
  SEmotion}, 2018.

\bibitem{destefanis2018measuring}
G.~Destefanis, M.~Ortu, D.~Bowes, M.~Marchesi, and R.~Tonelli, ``On measuring
  affects of github issues' commenters,'' in \emph{Proceedings of the 3rd
  International Workshop on SEmotion}, 2018.

\bibitem{ortu2018mining}
M.~Ortu, T.~Hall, M.~Marchesi, R.~Tonelli, D.~Bowes, and G.~Destefanis,
  ``Mining communication patterns in software development: A github analysis,''
  in \emph{Proceedings of the 14th PROMISE}, 2018.

\bibitem{venigalla2021understanding}
A.~S.~M. Venigalla and S.~Chimalakonda, ``Understanding emotions of developer
  community towards software documentation,'' in \emph{2021 IEEE/ACM 43rd ICSE:
  Software Engineering in Society (ICSE-SEIS)}.\hskip 1em plus 0.5em minus
  0.4em\relax IEEE, 2021.

\bibitem{neupane2019emod}
K.~P. Neupane, K.~Cheung, and Y.~Wang, ``Emod: An end-to-end approach for
  investigating emotion dynamics in software development,'' in \emph{2019 IEEE
  ICSME}.\hskip 1em plus 0.5em minus 0.4em\relax IEEE, 2019.

\bibitem{rong2022empirical}
S.~{Rong}, W.~{Wang}, U.~A. {Mannan}, E.~S. {de Almeida}, S.~{Zhou}, and
  I.~{Ahmed}, ``An empirical study of emoji use in software development
  communication,'' \emph{Information and Software Technology}, vol. 148, 2022.

\bibitem{tulili2023burnout}
T.~R. Tulili, A.~Capiluppi, and A.~Rastogi, ``Burnout in software engineering:
  A systematic mapping study,'' \emph{Information and Software Technology},
  vol. 155, 2023.

\bibitem{gao2023code}
S.~Gao, C.~Gao, Y.~He, J.~Zeng, L.~Nie, X.~Xia, and M.~Lyu, ``Code
  structure--guided transformer for source code summarization,'' \emph{ACM
  Transactions on Software Engineering and Methodology}, vol.~32, no.~1, 2023.

\bibitem{haque2022semantic}
S.~Haque, Z.~Eberhart, A.~Bansal, and C.~McMillan, ``Semantic similarity
  metrics for evaluating source code summarization,'' in \emph{Proceedings of
  the 30th IEEE/ACM International Conference on Program Comprehension}, 2022.

\bibitem{macneil2023experiences}
S.~MacNeil, A.~Tran, A.~Hellas, J.~Kim, S.~Sarsa, P.~Denny, S.~Bernstein, and
  J.~Leinonen, ``Experiences from using code explanations generated by large
  language models in a web software development e-book,'' in \emph{Proc. 54th
  ACM Tech. Symp. on CS Education V. 1}, 2023.

\bibitem{rai2022review}
S.~{Rai}, R.~C. {Belwal}, and A.~{Gupta}, ``A review on source code
  documentation,'' \emph{ACM Transactions on Intelligent Systems and
  Technology}, vol.~13, 2022.

\bibitem{ardimento2020using}
P.~Ardimento and C.~Mele, ``Using bert to predict bug-fixing time,'' in
  \emph{2020 IEEE Conference on Evolving and Adaptive Intelligent Systems
  (EAIS)}, 2020.

\bibitem{wang2023clebpi}
W.-Y. Wang, C.-H. Wu, and J.~He, ``Clebpi: Contrastive learning for bug
  priority inference,'' \emph{IST}, vol. 164, 2023.

\bibitem{ali2023bert}
A.~Ali, Y.~Xia, Q.~Umer, and M.~Osman, ``Bert based severity prediction of bug
  reports for the maintenance of mobile applications,'' \emph{JSS}, 2023.

\bibitem{keim2020does}
J.~Keim, A.~Kaplan, A.~Koziolek, and M.~Mirakhorli, ``Does bert understand
  code?--an exploratory study on the detection of architectural tactics in
  code,'' in \emph{European Conference on Software Architecture}.\hskip 1em
  plus 0.5em minus 0.4em\relax Springer, 2020.

\bibitem{sharma2022exploratory}
R.~Sharma, F.~Chen, F.~Fard, and D.~Lo, ``An exploratory study on code
  attention in bert,'' in \emph{Proceedings of the 30th IEEE/ACM ICPC}, 2022.

\bibitem{yang2022aspect}
C.~Yang, B.~Xu, J.~Y. Khan, G.~Uddin, D.~Han, Z.~Yang, and D.~Lo,
  ``Aspect-based api review classification: How far can pre-trained transformer
  model go?'' in \emph{2022 IEEE International Conference on SANER}.\hskip 1em
  plus 0.5em minus 0.4em\relax IEEE, 2022.

\bibitem{he2022ptm4tag}
J.~He, B.~Xu, Z.~Yang, D.~Han, C.~Yang, and D.~Lo, ``Ptm4tag: sharpening tag
  recommendation of stack overflow posts with pre-trained models,'' in
  \emph{Proceedings of the 30th IEEE/ACM ICPC}, 2022.

\end{thebibliography}

\end{sloppypar}

\end{document}